\documentclass[letter]{aa} 

\usepackage{graphicx}
\usepackage{txfonts}
\begin{document} 
\titlerunning{The long period of $^3$He-rich solar energetic particles}

   \title{The long period of $^3$He-rich solar energetic particles measured by Solar Orbiter on 2020 November 17--23}


   \author{R.~Bu\v{c}\'ik\inst{1}
          \and
          G.~M.~Mason\inst{2}
          \and
          R.~G\'omez-Herrero\inst{3}
          \and
          D.~Lario\inst{4}
          \and
          L.~Balmaceda\inst{4,5}
          \and
          N.~V.~Nitta\inst{6}
          \and
          V.~Krupa\v{r}\inst{4,7}
          \and
           N.~Dresing\inst{8,9}
           \and
           G.~C.~Ho\inst{2}
            \and
           R.~C.~Allen\inst{2}
            \and
           F.~Carcaboso\inst{3}
            \and
           J.~Rodr\'iguez-Pacheco\inst{3}
            \and
           F.~Schuller\inst{10}
            \and
           A.~Warmuth\inst{10}
            \and
           R.~F.~Wimmer-Schweingruber\inst{8}
            \and
           J. L. Freiherr von Forstner\inst{8,11}
            \and
           G. B. Andrews\inst{2}
            \and
            L. Berger\inst{8}
             \and
            I. Cernuda\inst{3}
             \and
            F. Espinosa Lara\inst{3}
             \and
            W. J. Lees\inst{2}
             \and
             C. Mart\'in\inst{8,12}
              \and
             D. Pacheco\inst{8}
              \and
             M. Prieto\inst{3}
              \and
             S. S\'anchez-Prieto\inst{3}
              \and
             C. E. Schlemm\inst{2}
              \and
              H. Seifert\inst{2}
               \and
              K.~Tyagi\inst{2,13}
              \and
              M.~Maksimovic\inst{14}
              \and
              A. Vecchio\inst{14,15}
              \and
              A. Kollhoff\inst{8}
              \and
              P. K\"{u}hl\inst{8}
              \and
              Z. G. Xu\inst{8}
              \and
              S. Eldrum\inst{8}
}

   \institute{Southwest Research Institute, San Antonio, TX 78238, USA\\
              \email{radoslav.bucik@swri.org}
         \and
            Applied Physics Laboratory, Johns Hopkins University, Laurel, MD 20723, USA
            \and
            Universidad de Alcal\'a, Space Research Group, 28805 Alcal\'a de Henares, Spain
            \and
            Heliophysics Science Division, NASA Goddard Space Flight Center, Greenbelt, MD, USA 
            \and
            George Mason University, Fairfax, VA, USA 
             \and
Lockheed Martin Advanced Technology Center, Palo Alto, CA 94304, USA
   \and
Goddard Planetary Heliophysics Institute, University of Maryland, Baltimore County, Baltimore, MD 21250, USA 
   \and
Institut f\"{u}r Experimentelle und Angewandte Physik, Christian-Albrechts-Universit\"{a}t zu Kiel, Kiel, Germany 
   \and
now at Department of Physics and Astronomy, University of Turku, FI-20014 Turku, Finland 
   \and
Leibniz-Institut f\"{u}r Astrophysik Potsdam, D-14482 Potsdam, Germany
   \and
 now at Paradox Cat GmbH, 80333 M\"{u}nchen, Germany
   \and
now at German Aerospace Center (DLR), Berlin, Germany, Dept. of Extrasolar Planets and Atmospheres
   \and
now at Univ. Colorado/LASP, Boulder, CO, USA
\and
LESIA, Observatoire de Paris, Universit\'e PSL, CNRS, Sorbonne Universit\'e, Universit\'e de Paris, France
\and
Radboud Radio Lab, Department of Astrophysics, Radboud University Nijmegen, The Netherlands
             }

   \date{Received ; accepted }

 
  \abstract{

We report observations of a relatively long period of $^3$He-rich solar energetic particles (SEPs) measured by Solar Orbiter. The period consists of several well-resolved ion injections. The high-resolution STEREO-A imaging observations reveal that the injections coincide with EUV jets/brightenings near the east limb, not far from the nominal magnetic connection of Solar Orbiter. The jets originated in two adjacent, large, and complex active regions as observed by the Solar Dynamics Observatory when the regions rotated to the Earth’s view. It appears that the sustained ion injections were related to the complex configuration of the sunspot group and the long period of $^3$He-rich SEPs to the longitudinal extent covered by the group during the analyzed time period. 

}

   \maketitle
%

\section{Introduction}

$^3$He-rich solar energetic particle (SEP) events show enormous enhancements of rare species such as the nuclide $^3$He and ultra-heavy elements by factors up to $\sim$10$^4$ above the nominal coronal abundances \citep[e.g.,][]{2007SSRv..130..231M,2021LNP...978.....R}. The events are highly associated ($>$95\%) with type III radio bursts \cite[e.g.,][]{1986ApJ...308..902R,2006ApJ...650..438N}, the emission generated by $\sim$10--100\,keV outward streaming electrons. Solar sources of $^3$He-rich SEPs have been associated with EUV jets \citep[][and references therein]{2020SSRv..216...24B}, suggesting acceleration via magnetic reconnection involving field lines open to interplanetary space \citep{2002ApJ...571L..63R}. Progress in understanding $^3$He-rich SEPs has been hampered by the low intensities and short duration of these events. 
Solar Orbiter \citep{2020A&A...642A...1M} will enable unprecedented studies of small-size $^3$He-rich SEP events combining in-situ and remote-sensing observations close to the Sun.

The first Solar Orbiter $^3$He-rich SEP events were measured during the spacecraft’s first perihelion pass from 0.52 to 0.96\,au \citep{2020A&AM} in June--September 2020. Three events out of the five discrete events reported by \citet{2020A&AM} have a 0.2--2\, MeV/nucleon $^3$He/$^4$He ratio above 10\% with a maximum $^3$He/$^4$He ratio of 0.61. In this paper, we report a relatively long period of $^3$He-rich SEPs, spanning almost 7 days in November 2020, observed by Solar Orbiter near 0.9\,au. Such a long period may indicate a nearly continuous $^3$He-rich SEP injection into the interplanetary space \citep{2007SSRv..130..231M}.

\section{Observations}

The $^3$He-rich SEPs reported in this paper were measured by the Suprathermal Ion Spectrograph (SIS) of the Energetic Particle Detector (EPD) suite \citep{2020A&A...642A...7R} aboard Solar Orbiter. SIS is a time-of-flight mass spectrometer that measures elemental composition from H through ultra-heavy nuclei in the kinetic energy range of $\sim$0.1--10\,MeV/nucleon. SIS has two telescopes, one pointing 30$^\circ$ (sunward) and the other 160$^\circ$ (anti-sunward) to the west of the spacecraft-Sun line. We also use energetic electron measurements made by the Electron Proton Telescope (EPT) of EPD, which covers energies (20--400\,keV) between two other instruments of the EPD suite, STEP, and HET. The first year of operations, and details of the data products, provided by EPD, can be found in \citet{2020A&AW}.

Solar sources of $^3$He-rich SEPs were examined using high-resolution EUV images from the SECCHI/EUVI instrument \citep{2008SSRv..136...67H} on the STEREO-A (STA). The EUVI provides full-disk images of the Sun with 3$\arcsec$ spatial and 5-minute nominal temporal resolution in four wavelength channels (304, 171, 195, and 284\,{\AA}). We use the 195\,{\AA} images that have the highest temporal resolution (5.0 and 2.5 minutes) in the examined period. The Extreme-Ultraviolet Imager \citep[EUI;][]{2020A&A...642A...8R} on Solar Orbiter provides images only with limited spatial and temporal resolution during the aforementioned period. We note that until November 2021, Solar Orbiter is in the cruise phase when remote-sensing instruments are only occasionally switched on for calibration. Further, we inspect radio spectrograms for the presence of the associated type III radio bursts. The radio data are provided by the Solar Orbiter Radio and Plasma Waves \citep[RPW;][]{2020A&A...642A..12M} and the STEREO-A Waves \citep{2008SSRv..136..487B} instruments with a frequency range ($<$16\,MHz) covering emission generated from about $\sim$2\,R$_\sun$ to 1\,au. We also make use of full-disk line-of-sight magnetograms obtained from Helioseismic and Magnetic Imager \citep[HMI;][]{2012SoPh..275..207S} onboard Solar Dynamics Observatory (SDO).

\begin{figure}
   \centering
   \includegraphics[width=7cm]{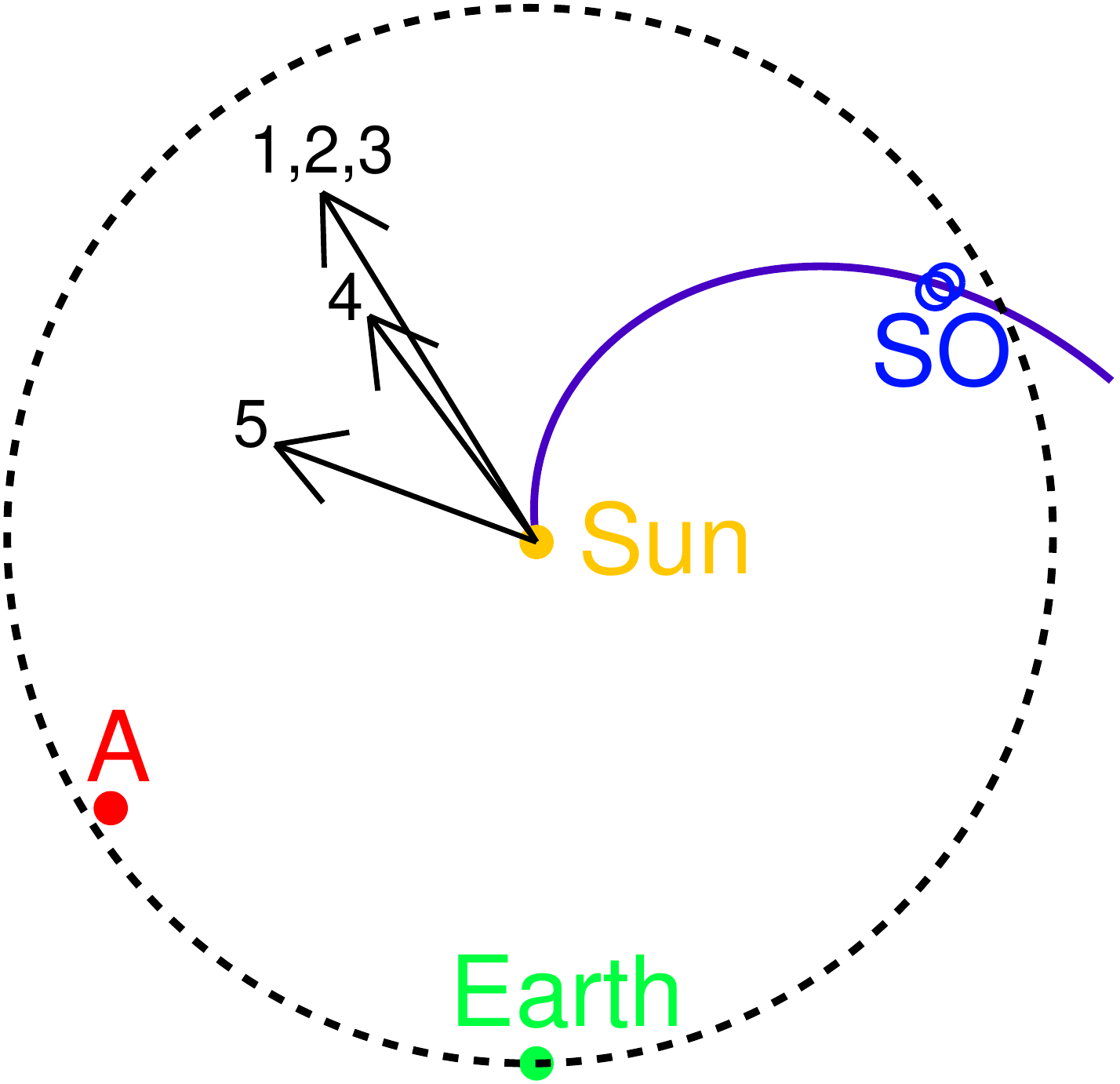}
   \caption{Ecliptic plane projection of the Solar Orbiter (SO), STEREO-A (A), and the Earth in 2020 November 17--23. Two overlapping rings mark SO at the beginning and end of the examined period. The arrows indicate the longitude of the solar source associated with ion injections. Parker spiral for 340\,km/s solar wind speed connecting to Solar Orbiter is shown. The dashed line corresponds to 1\,au orbit.}
              \label{orbit}%
    \end{figure}

The location of Solar Orbiter and STEREO-A during the investigated period is shown in Fig.~\ref{orbit}. Solar Orbiter traveled from 0.93 to 0.91\,au; STEREO-A remained at 0.96\,au. Both spacecraft were near the ecliptic plane, Solar Orbiter at $-$6$^{\circ}$ and STEREO-A at 7$^{\circ}$ of heliographic latitude. The angular separation between Solar Orbiter and STEREO-A was 180$^{\circ}$. SDO is in orbit around the Earth.

%

   \begin{figure}
   \centering
\hspace*{-0.2cm}  
     \includegraphics[width=9.2cm]{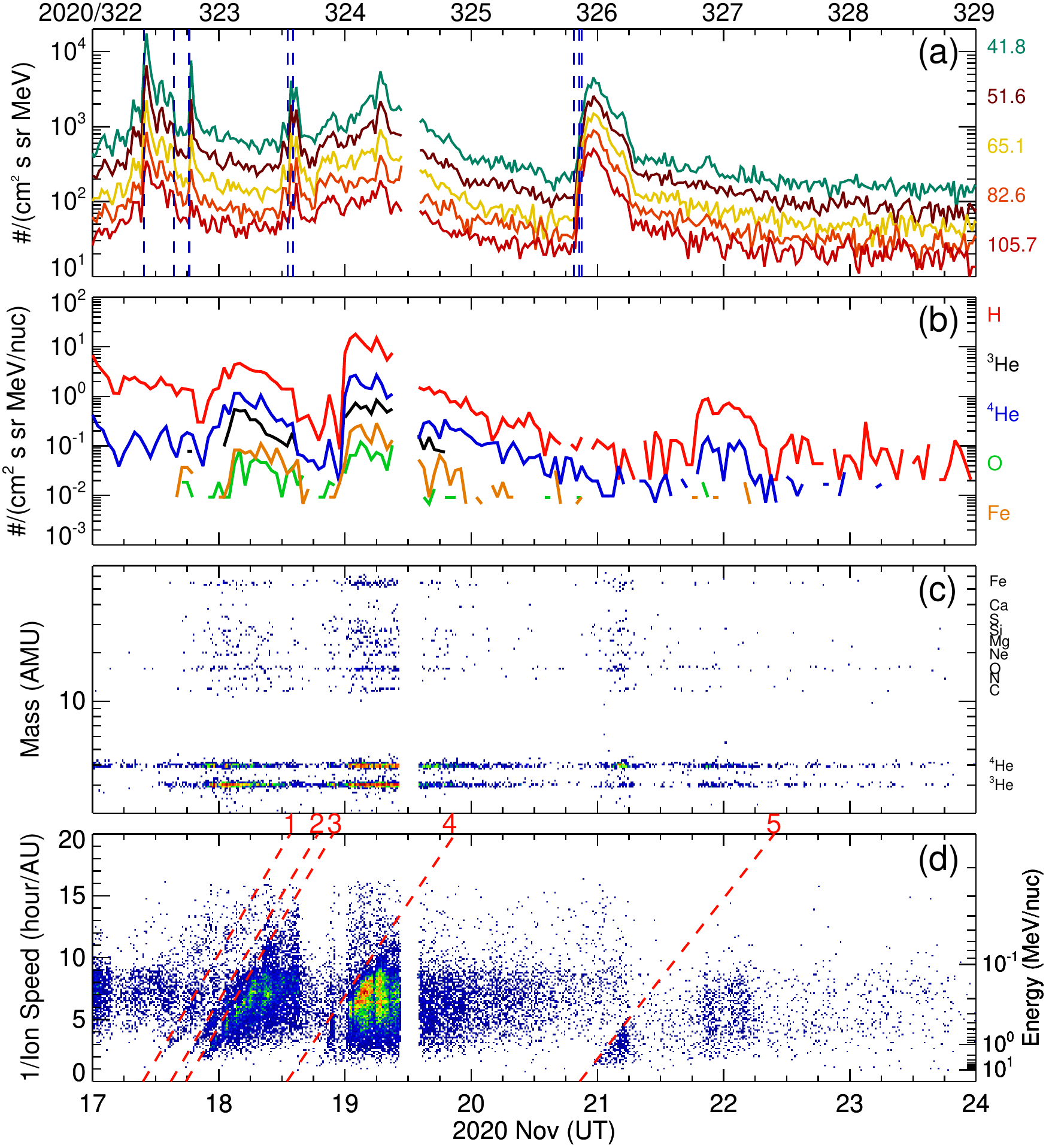}
   \caption{(a) 30-minute electron intensities from EPT (41.8--105.7\,keV) sunward pointing sensor. Dashed vertical lines mark type III radio bursts listed in Table~\ref{t1}. (b) SIS 1-hr H, $^3$He, $^4$He, O, and Fe intensities at 0.23--0.32\,MeV/nucleon. (c) SIS  mass spectrogram at 0.4--10\,MeV/nucleon. (d) SIS 1/speed vs. arrival times of 2--70\,AMU ions. Sloped dashed lines approximately mark the ion injections. SIS measurements are from both telescopes averaged together. The gap around the noon of November 19 was caused by EPD being shut down for software maintenance.}
              \label{panels}%
    \end{figure}

\begin{figure}
   \centering
        \includegraphics[width=0.45\textwidth]{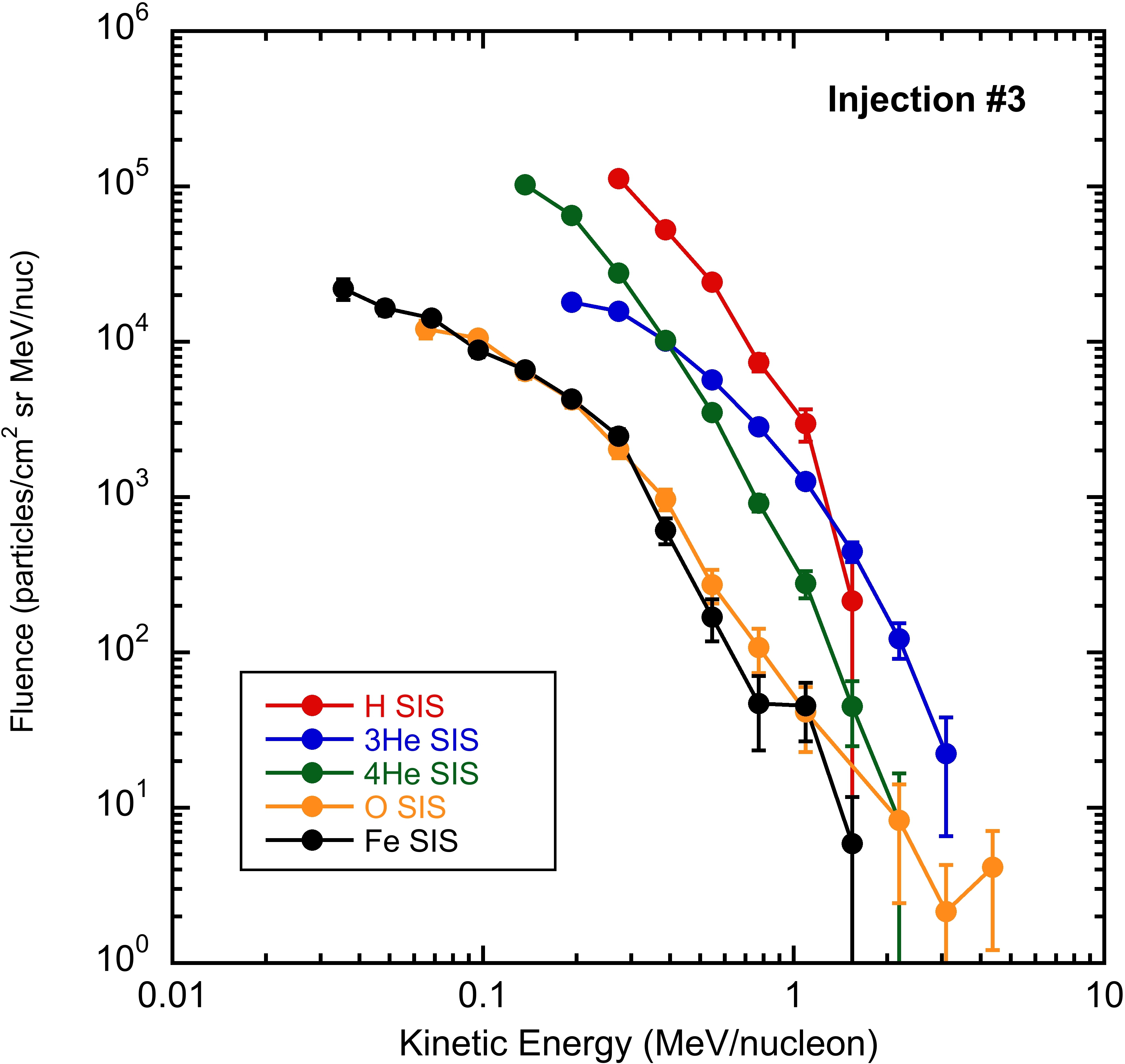}
   \caption{Fluence spectra for selected species in injection \#3.}
              \label{spec}%
    \end{figure}

Figure~\ref{panels} displays Solar Orbiter EPT and SIS measurements in 2020 November 17--23. Figure~\ref{panels}a presents 30-minute electron intensities at different energy bins between 41.8 and 105.7\,keV. Figure~\ref{panels}b shows hourly averages of the 0.23--0.32 MeV/nucleon H, $^3$He, $^4$He, O and Fe intensities as measured by both telescopes of SIS. Three major increases are seen at the intensity time profiles that start near the end of November 17, near the end of November 18, and around midday of November 21. It is clearly seen that the first two increases are $^3$He- and Fe-rich. The mass spectrogram in Fig.~\ref{panels}\c shows almost continuous $^3$He presence from the middle of November 17 through the end of November 23, i.e., $\sim$6.5 days. The inverse ion-velocity time spectrogram in Fig.~\ref{panels}d shows at least three ion injections contributing to the first increase on November 17--18; one injection contributing to the increase on November 18--20 and one ion injection for the increase on November 21. These injections can be identified based on their characteristic triangular pattern in the inverse speed plots. Figure~\ref{spec} shows the fluence energy spectra for selected ion species in injection \#3 where enhancements in all ion species were observed without the inconvenient data gaps as in the case of injection \#4. The $^3$He, O, and Fe show rollovers toward low energies as it has been previously reported in many $^3$He-rich SEP events \citep[e.g.,][]{2000ApJ...545L.157M}. The ion fluences for injections \#1, \#2, and \#3 are integrated in swoosh boxes bounded by slanted lines 1 \& 2, 2 \& 3, and 3 \& 4, respectively. For injection \#4 the swoosh box is between the slanted line 4 and November 20 19:26\,UT and for injection \#5 between slanted line 5 and November 22 08:24\,UT.

\begin{table*}
\caption{\label{t1}Characteristics of the $^3$He-rich period. See text for more details.}
\centering
\begin{tabular}{c c c c c c c c c}
\hline\hline
 &Ion injection&Type III&\multicolumn{2}{c}{STA EUVI event}&Separation&Elec. injection&$^3$He/$^4$He\tablefootmark{c}&Fe/O\tablefootmark{c}\\
\cline{4-5}
 &time (UT)&start (UT)&Type\tablefootmark{a}&Location&angle\tablefootmark{b}($^\circ$)&time (UT)&&\\
\hline
1&322.42  Nov-17 10:05&09:49 [41]&B&E90S22&20&09:20&0.61$\pm$0.08&2.00$\pm$0.37\\
2&322.62  Nov-17 14:53&15:28 [20]&B&E90S18&20&...&0.22$\pm$0.03&0.63$\pm$0.06\\
3&322.74  Nov-17 17:46&18:20 [12]&J&E90S18&20&18:20&0.90$\pm$0.03&0.91$\pm$0.01\\
  &		                      &18:24 [16]&…&E90S18&20&…&…&…\\
 4&323.54  Nov-18 12:58&13:08 [00]	&B&E85S23&25&13:10&	0.56$\pm$0.01&1.35$\pm$0.01\\
  &                                    &14:09 [01]&B&E85S23&25&14:07&…&…\\
 5&325.87  Nov-20 20:53&19:34 [26]	&B&E48S19&62&19:30&	0.32$\pm$0.03&0.76$\pm$0.03\\
   &                                   &20:33 [25]&J&E52S17&58&…&…&…\\
   &		                       &21:00 [52]&B&E48S19&62&…&…&…\\
\hline
\end{tabular}
\tablefoot{
\tablefoottext{a}{B: brightening; J: jet}
\tablefoottext{b}{between Solar Orbiter magnetic footpoint longitude on the Sun and the longitude of the EUVI event}
\tablefoottext{c}{0.2--2.0\,MeV/nucleon}
}
\end{table*}

\begin{figure*}
   \centering
          \includegraphics[width=0.8\textwidth]{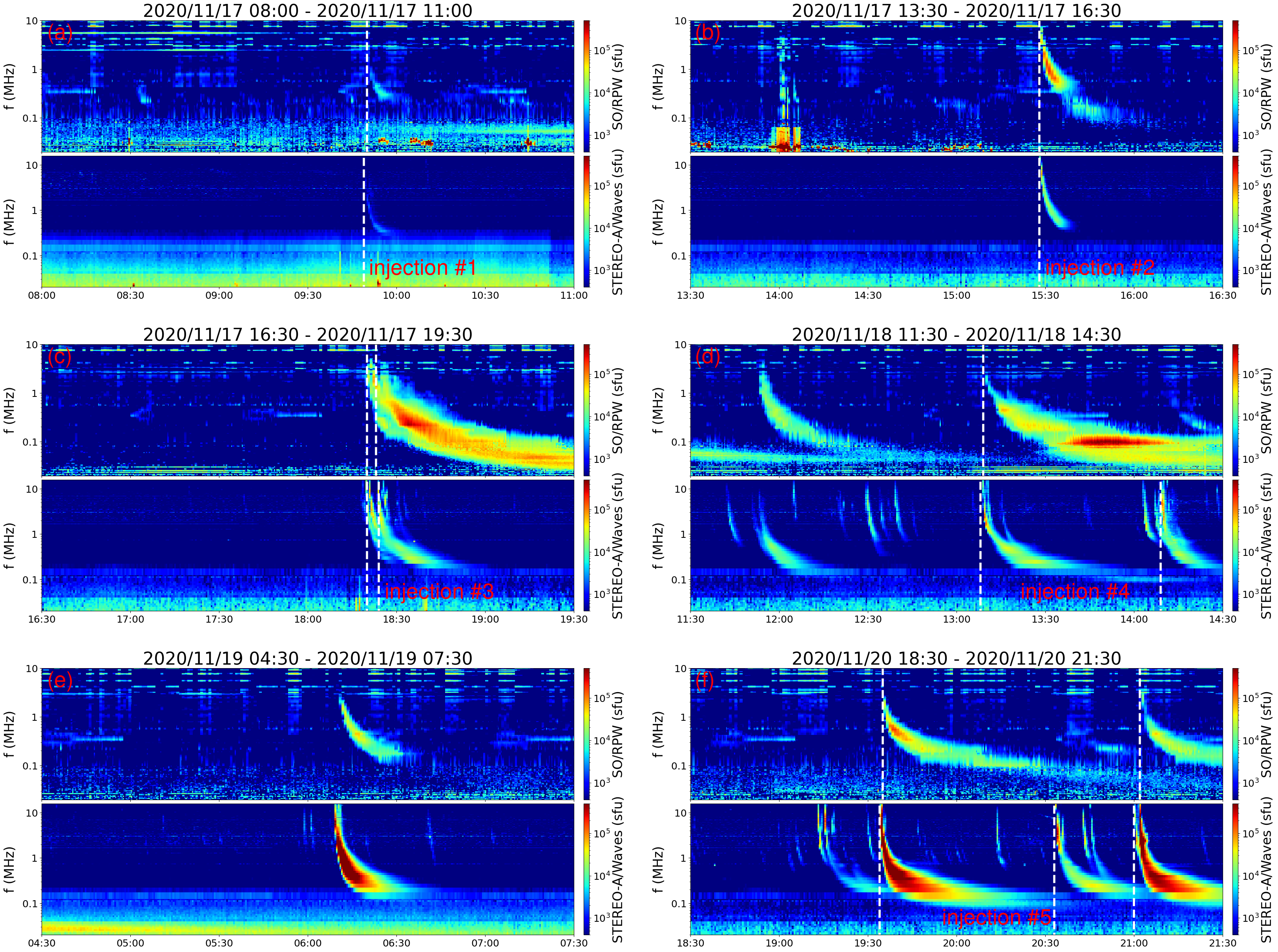}
   \caption{(a--f) Solar Orbiter/RPW and STEREO-A/Waves radio spectrograms. (a--d) and (f) correspond to ion injections \#1 -- \#4 and \#5, respectively. (e) corresponds to an electron event. The vertical dashed lines mark the start times of type III radio bursts associated with the ion injections.}
              \label{radio}%
    \end{figure*}

Table~\ref{t1} lists the characteristics of the $^3$He-rich period. Column 1 indicates the injection number, column 2 the ion injection time at the Sun (as a day of year and date), estimated by extrapolation of dispersive signature of individual ions in the inverted velocity-time spectrogram indicated by the inclined dashed red lines in Fig.~\ref{panels}d. The uncertainty in the injection time estimated by this method is $\pm$45\,min \citep{2000ApJ...545L.157M,2016A&A...585A.119W}. Column 3 gives the associated type III radio burst start times as observed by STEREO-A/Waves at 16\,MHz. Note, the RPW showed an enhanced level of interference at higher frequencies. Multiple type III bursts appear to contribute to injections \#3, \#4, and \#5 (see Fig.~\ref{radio}). To compare with the estimated ion injection times, the square brackets show minutes of the type III bursts start times after subtraction of the light travel time ($\sim$8\,min). It is unclear if the type III burst at 19:34\,UT is associated with injection \#5; it occurs too early to be within $\pm$45\,min error of ion estimated release time. We note that injection \#5 is the weakest of all the injections and the magnetic connection to the site was interrupted early in the event, around $\sim$08:00 on November 21 (Fig.~\ref{panels}d), as indicated by the abrupt drop in ion counts at all energies. Therefore, the estimated injection time is only tentative, and the association with the type III at 19:34\,UT could not be ruled out. Columns 4 and 5 provide the type and location of the associated parent solar eruption, respectively, as seen in the EUVI on STEREO-A, where J indicates a clear EUV jet moving away from the parent active region and B indicates just a brightening seen in the EUV images without apparent outward movement. We cannot identify the type of EUVI event in 5-minute resolution images for the 2nd type III burst, corresponding to injection \#3, that occurred only 4 minutes after the previous type III burst. Column 6 indicates a separation angle between Solar Orbiter magnetic footpoint longitude on the Sun and the longitude of the EUVI event. The magnetic footpoint of Solar Orbiter, based on simple Parker spiral approximation and assuming solar wind speed of 340\,km/s, was $\sim$W70 which corresponds to E110 from STEREO-A view. The value of 340\,km/s is the median solar wind speed measured by SWEPAM \citep{1998SSRv...86..563M} on ACE nine days earlier (November 8--12), which corresponds to the solar rotation between the L1 and Solar Orbiter separated by 122$^{\circ}$. The Solar Orbiter Solar Wind Analyser \citep[SWA;][]{2020A&A...642A..16O} data were not available for the examined period. Column 7 shows the electron injection time, estimated from the inverted velocity-time spectrogram (not shown) of 1-min averaged EPD electron data. Columns 8 and 9 provide $^3$He/$^4$He and Fe/O ratios at 0.2--2.0\,MeV/nucleon.

Figure~\ref{radio} shows Solar Orbiter and STEREO-A radio spectrograms where we have indicated the type III radio bursts associated with the ion injections. The presence of high frequencies at Waves in all the bursts suggests that the source was not behind the east limb as seen from STEREO-A. The 2nd type III bursts in injections \#4 and \#5 were weak at Solar Orbiter. During ion injection \#3, two small dispersive electron events were detected by EPD with solar injections on November 18 10:25, and 11:45\,UT. The later one appears as a small peak in the EPT intensity-time profile at $\sim$12:00\,UT on November 18 (Fig.~\ref{panels}a). Figure~\ref{radio}d shows a type III burst associated with the electron injection at 11:45\,UT. The type III burst associated with the electron injection at 10:25\,UT is clearly observed only by Solar Orbiter (see Fig.~\ref{radio}d for low-frequency part, $\sim$0.05\,MHz, between 11:30 and $\sim$13:00\,UT). During ion injection \#4, another small dispersive electron event was measured with solar injection on November 19 06:00\,UT (see Fig.~\ref{panels}a for the peak at $\sim$07:00\,UT on November 19). Figure~\ref{radio}e shows the associated type III radio burst. The type III bursts related to these electron events were accompanied by EUV jets (see Fig.~\ref{euv4}). 

\begin{figure}
   \centering
   \includegraphics[width=9cm]{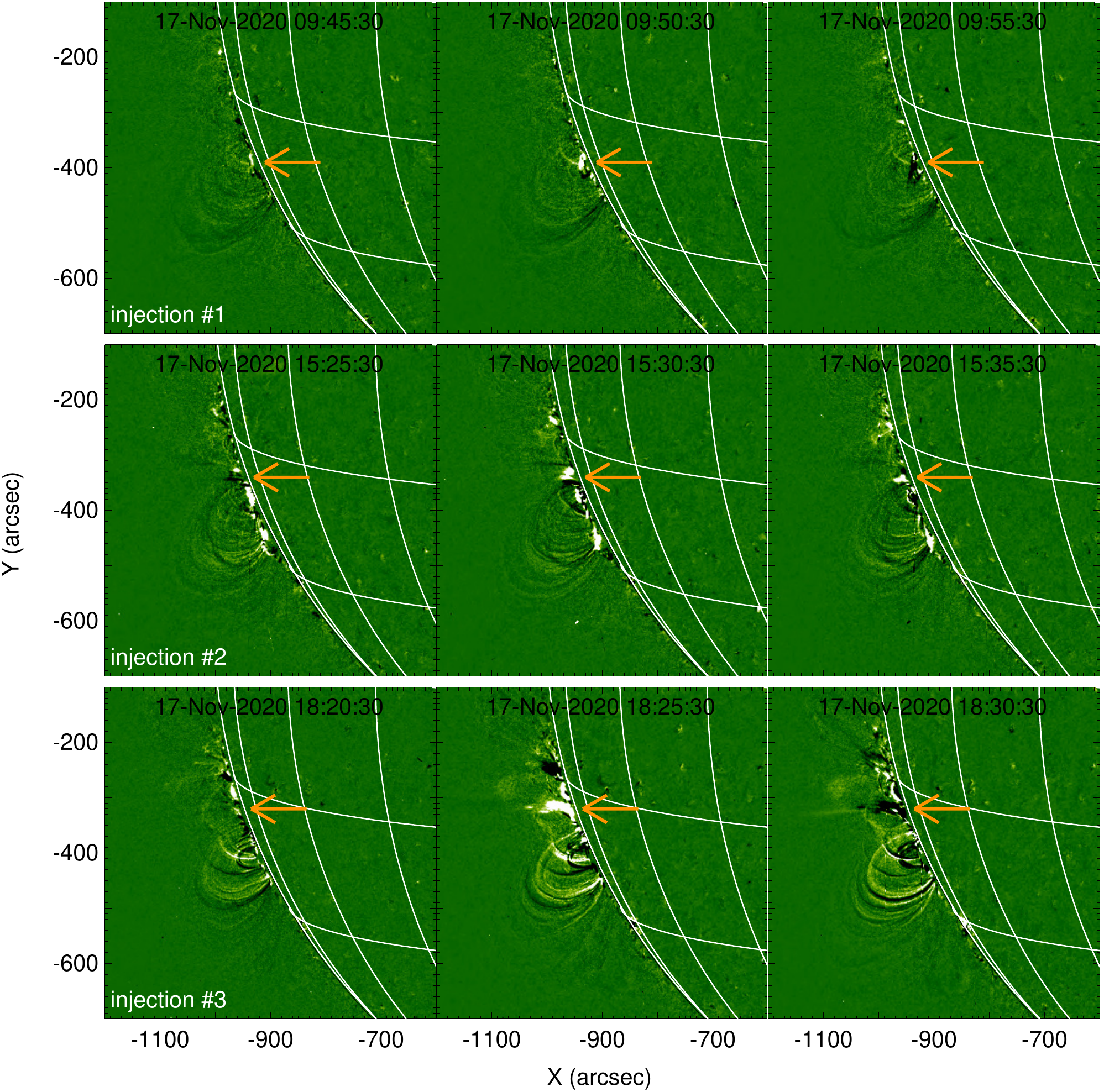}
   \caption{STEREO-A 195\,{\AA} EUV running difference images corresponding to injection \#1 (top row), \#2 (middle row), and \#3 (bottom row). The arrow marks the solar source. The heliographic longitude-latitude grid has a 15$^\circ$ spacing.}
              \label{euv1}%
    \end{figure}

To identify solar sources, we inspect full-disk solar images for EUV brightenings as seen by STEREO-A that temporally coincide with the type III radio burst associated with the ion injection. Figure~\ref{euv1} shows the EUV activity around the times of the type III radio burst for injection \#1 (top row), \#2 (middle row), and \#3 (bottom row). We do not see clear jets for injections \#1 and \#2 in EUVI images. The EUV images of the solar source for injections \#4 and \#5 are shown in Appendix~A. We note that SDO was not well located to observe EUV activity related to the origin of these ion injections (Fig.~\ref{orbit}). However, for injection \#5 the SDO Atmospheric Imaging Assembly  
\citep[AIA;][]{2012SoPh..275...17L} observed the EUV jets, from region $\sim$16$^\circ$ behind the east limb, that temporally match all three type III bursts.

On November 17--18, the EUI on Solar Orbiter provides images only with 1-hr cadence. On November 19--20, there are also higher cadence data, but they either cover only short periods or they have a low spatial resolution. In any case, periods of jets/brightening on November 19 and 20 were missed by EUI. 

   \begin{figure*}
   \centering
     \includegraphics[width=0.8\textwidth]{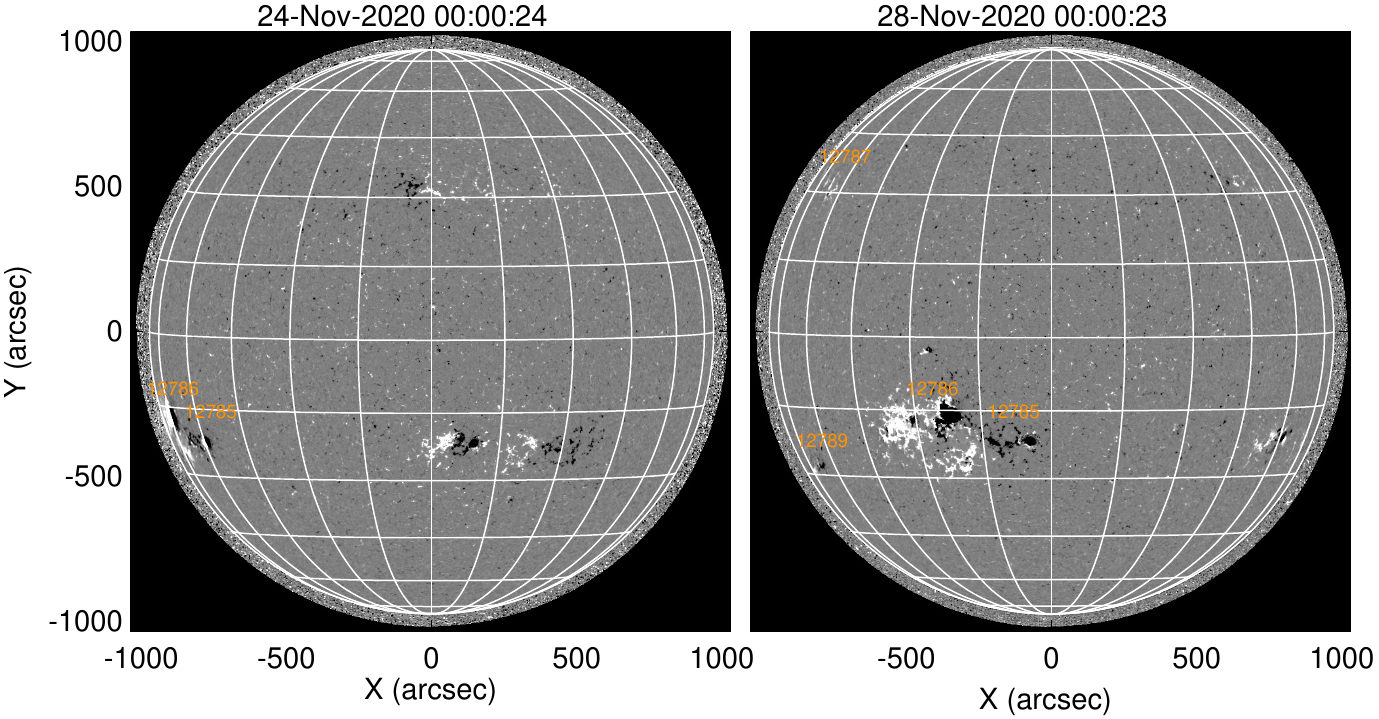}
   \caption{SDO HMI line-of-sight magnetograms (scaled to $\pm$100\,G). The numbers mark the NOAA active regions of interest. The heliographic longitude-latitude grid has a 15$^\circ$ spacing.}
              \label{hmi}%
    \end{figure*}

Figure~\ref{hmi} shows SDO HMI magnetograms on November 24 00:00 UT (Left) and November 28 00:00 UT (Right). The EUV activity observed by STEREO-A on November 17-20 likely originated in two adjacent, large active regions (ARs) 12785 and 12786 that appeared near the east limb, as viewed from Earth early on November 23. The latitude of the jets/brightenings as indicated in Table~\ref{t1} and as seen in Fig.~\ref{euv1} and Fig.~\ref{euv2}--\ref{euv4} matches well with the latitudes of these two ARs. It is particularly well seen in Fig.~\ref{euv3}, where the constellations of bright areas are similar to the positions of these two ARs. Thus, the brightenings in injection \#5 can be clearly associated with AR 12785, while the jet occurred between these ARs. The AR 12786 shows a complex $\beta\gamma$ magnetic class\footnote{$\beta\gamma$ denotes a bipolar sunspot group with no clearly marked line separating spots of opposite polarity; $\beta$ indicates a bipolar sunspot group} and sunspot area of 1000 millionths of the solar hemisphere (MH) on November 25--26. The AR 12785 has a simple $\beta$ magnetic class and sunspot area of 140 MH (November 23--24). This information is provided by the USAF/NOAA Solar Region Summary (ftp.swpc.noaa.gov/pub/warehouse/2020/SRS). The magnetic complexity of the ARs decreased after they crossed the central meridian (as observed by SDO) on November 29--30. These ARs were seen by STEREO-A in EUV for the first time on November 19 (they were not reported in the previous rotation), and therefore we do not know what their properties were on November 17. As STEREO-A does not have a magnetograph, the magnetic class and area of the ARs were unknown when the examined activity was occurring. Marked are also two small ARs 12787 and 12789 (Fig.~\ref{hmi}  Right) that could be located close to the Solar Orbiter nominal magnetic footpoint longitude. If these regions were in the hidden hemisphere, we cannot confirm/rule out that there was some simultaneous activity occurring in them as well. However, it is improbable that these regions dominated the observed long period of $^3$He-rich SEPs as all type III radio bursts temporally coincide with jets/brightening in AR 12786 and 12785. 

\section{Discussion and conclusion}
The relatively long period of $^3$He-rich SEPs observed by Solar Orbiter is related to the recurrent activity (brightening and jets) in a large and complex group of sunspots in two adjacent ARs. Recurrent $^3$He-rich SEP events have been found to originate from active regions at the boundary of low-latitude coronal holes \citep[e.g.,][]{2006ApJ...639..495W,2014ApJ...786...71B}. There are only a few reports of $^3$He-rich SEPs associated with sunspot jets \citep{2008ApJ...675L.125N,2018ApJ...869L..21B} and none report recurrent ion injection. The configuration with two large and complex nearby ARs may be favorable for the recurrent particle injections in the sense that there may be long-lived interaction between the negative polarity of one AR and the positive polarity of neighboring AR leading to the magnetic reconnection. Furthermore, these two ARs produce a longitudinally extended source ($\sim$40$^\circ$) where spacecraft may be magnetically connected for a long period as the Sun rotates. We note that this extended region is rotating away from Solar Orbiter such that the magnetic connection is presumably weakening with time.

\citet{2008ApJS..176..497K} have studied extended periods of $^3$He-rich SEPs where most of them showed no dispersive onset. The authors have suggested that the temporal confinement of ions in the solar wind structures is an essential factor in the occurrence of such periods. \citet{2015AA...580A..16C} have reported two relatively long, 4-day, periods of $^3$He-rich SEPs that were produced by recurring injections originating in plage regions from dispersed sunspots. While \citet{2015AA...580A..16C} have identified two injections per period, we report at least five ion injections responsible for a long period. There might be other unresolved ion injections during the decay phase of the 1st and the 2nd ion intensity increases.

The recurrent production of $^3$He-rich SEPs appears to occur in different magnetic environments that include plages, coronal holes, and sunspots and may be the result of a common process. Further studies may confirm whether complex and longitudinally extended sunspot groups are responsible for longer $^3$He-rich SEP periods compared to simple and small size sources.

\begin{acknowledgements}
 The Suprathermal Ion Spectrograph (SIS) is a European facility instrument funded by ESA.  The SIS instrument was constructed by the JHU/Applied Physics Lab. and CAU Kiel.  We thank the many individuals at ESA and within the Energetic Particle Detector team for their support in its development.  Post launch operation of SIS at APL is funded by NASA contract NNN06AA01C, and we thank NASA headquarters and the NASA/GSFC Solar Orbiter project office for their continuing support.  The UAH team acknowledges the financial support by the Spanish Ministerio de Ciencia, Innovaci\'on y Universidades FEDER/MCIU/AEI Projects ESP2017-88436-R and PID2019-104863RB- I00/AEI/10.13039/501100011033. The CAU Kiel team thanks the German Federal Ministry for Economic Affairs and Energy and the German Space Agency (Deutsches Zentrum f\"ur Luft- und Raumfahrt, e.V., (DLR)) for their unwavering support under grant numbers 50OT0901, 50OT1202, 50OT1702, and 50OT2002; and ESA for supporting the build of SIS under contract number SOL.ASTR.CON.00004, and the University of Kiel and the Land Schleswig-Holstein for their support of SIS. F. Carcaboso acknowledges the financial support by the Spanish MINECO-FPI-2016 predoctoral grant with FSE. V. Krupa\v{r} acknowledges the support by NASA under grants 18-2HSWO2182-0010 and 19-HSR-192-0143.  The work was partly supported by NASA grant 80NSSC19K0079. LB acknowledges the support from the NASA program NNH17ZDA001N-LWS (Awards Nr. 80NSSC19K1261 and 80NSSC19K1235).
\end{acknowledgements}

%
   \bibliographystyle{aa} 
   \bibliography{ads} 

\begin{thebibliography}{26}
\expandafter\ifx\csname natexlab\endcsname\relax\def\natexlab#1{#1}\fi

\bibitem[{{Bougeret} {et~al.}(2008){Bougeret}, {Goetz}, {Kaiser}, {Bale},
  {Kellogg}, {Maksimovic}, {Monge}, {Monson}, {Astier}, {Davy}, {Dekkali},
  {Hinze}, {Manning}, {Aguilar-Rodriguez}, {Bonnin}, {Briand}, {Cairns},
  {Cattell}, {Cecconi}, {Eastwood}, {Ergun}, {Fainberg}, {Hoang}, {Huttunen},
  {Krucker}, {Lecacheux}, {MacDowall}, {Macher}, {Mangeney}, {Meetre},
  {Moussas}, {Nguyen}, {Oswald}, {Pulupa}, {Reiner}, {Robinson}, {Rucker},
  {Salem}, {Santolik}, {Silvis}, {Ullrich}, {Zarka}, \&
  {Zouganelis}}]{2008SSRv..136..487B}
{Bougeret}, J.~L., {Goetz}, K., {Kaiser}, M.~L., {et~al.} 2008, \ssr, 136, 487

\bibitem[{{Bu{\v{c}}{\'\i}k}(2020)}]{2020SSRv..216...24B}
{Bu{\v{c}}{\'\i}k}, R. 2020, \ssr, 216, 24

\bibitem[{{Bu{\v{c}}{\'\i}k} {et~al.}(2014){Bu{\v{c}}{\'\i}k}, {Innes}, {Mall},
  {Korth}, {Mason}, \& {G{\'o}mez-Herrero}}]{2014ApJ...786...71B}
{Bu{\v{c}}{\'\i}k}, R., {Innes}, D.~E., {Mall}, U., {et~al.} 2014, Astrophys.
  J., 786, 71

\bibitem[{{Bu{\v{c}}{\'\i}k} {et~al.}(2018){Bu{\v{c}}{\'\i}k}, {Wiedenbeck},
  {Mason}, {G{\'o}mez-Herrero}, {Nitta}, \& {Wang}}]{2018ApJ...869L..21B}
{Bu{\v{c}}{\'\i}k}, R., {Wiedenbeck}, M.~E., {Mason}, G.~M., {et~al.} 2018,
  Astrophys. J. Lett., 869, L21

\bibitem[{{Chen} {et~al.}(2015){Chen}, {Bu{\v{c}}{\'\i}k}, {Innes}, \&
  {Mason}}]{2015AA...580A..16C}
{Chen}, N.-H., {Bu{\v{c}}{\'\i}k}, R., {Innes}, D.~E., \& {Mason}, G.~M. 2015,
  Astron. Astrophys., 580, A16

\bibitem[{{Howard} {et~al.}(2008){Howard}, {Moses}, {Vourlidas}, {Newmark},
  {Socker}, {Plunkett}, {Korendyke}, {Cook}, {Hurley}, {Davila}, {Thompson},
  {St Cyr}, {Mentzell}, {Mehalick}, {Lemen}, {Wuelser}, {Duncan}, {Tarbell},
  {Wolfson}, {Moore}, {Harrison}, {Waltham}, {Lang}, {Davis}, {Eyles},
  {Mapson-Menard}, {Simnett}, {Halain}, {Defise}, {Mazy}, {Rochus}, {Mercier},
  {Ravet}, {Delmotte}, {Auchere}, {Delaboudiniere}, {Bothmer}, {Deutsch},
  {Wang}, {Rich}, {Cooper}, {Stephens}, {Maahs}, {Baugh}, {McMullin}, \&
  {Carter}}]{2008SSRv..136...67H}
{Howard}, R.~A., {Moses}, J.~D., {Vourlidas}, A., {et~al.} 2008, \ssr, 136, 67

\bibitem[{{Kocharov} {et~al.}(2008){Kocharov}, {Laivola}, {Mason}, {Didkovsky},
  \& {Judge}}]{2008ApJS..176..497K}
{Kocharov}, L., {Laivola}, J., {Mason}, G.~M., {Didkovsky}, L., \& {Judge},
  D.~L. 2008, Astrophys. J. Suppl. Ser., 176, 497

\bibitem[{{Lemen} {et~al.}(2012){Lemen}, {Title}, {Akin}, {Boerner}, {Chou},
  {Drake}, {Duncan}, {Edwards}, {Friedlaender}, {Heyman}, {Hurlburt}, {Katz},
  {Kushner}, {Levay}, {Lindgren}, {Mathur}, {McFeaters}, {Mitchell}, {Rehse},
  {Schrijver}, {Springer}, {Stern}, {Tarbell}, {Wuelser}, {Wolfson}, {Yanari},
  {Bookbinder}, {Cheimets}, {Caldwell}, {Deluca}, {Gates}, {Golub}, {Park},
  {Podgorski}, {Bush}, {Scherrer}, {Gummin}, {Smith}, {Auker}, {Jerram},
  {Pool}, {Soufli}, {Windt}, {Beardsley}, {Clapp}, {Lang}, \&
  {Waltham}}]{2012SoPh..275...17L}
{Lemen}, J.~R., {Title}, A.~M., {Akin}, D.~J., {et~al.} 2012, \solphys, 275, 17

\bibitem[{{Maksimovic} {et~al.}(2020){Maksimovic}, {Bale}, {Chust},
  {Khotyaintsev}, {Krasnoselskikh}, {Kretzschmar}, {Plettemeier}, {Rucker},
  {Sou{\v{c}}ek}, {Steller}, {{\v{S}}tver{\'a}k}, {Tr{\'a}vn{\'\i}{\v{c}}ek},
  {Vaivads}, {Chaintreuil}, {Dekkali}, {Alexandrova}, {Astier}, {Barbary},
  {B{\'e}rard}, {Bonnin}, {Boughedada}, {Cecconi}, {Chapron}, {Chariet},
  {Collin}, {de Conchy}, {Dias}, {Gu{\'e}guen}, {Lamy}, {Leray}, {Lion},
  {Malac-Allain}, {Matteini}, {Nguyen}, {Pantellini}, {Parisot}, {Plasson},
  {Thijs}, {Vecchio}, {Fratter}, {Bellouard}, {Lorf{\`e}vre}, {Danto},
  {Julien}, {Guilhem}, {Fiachetti}, {Sanisidro}, {Laffaye}, {Gonzalez},
  {Pontet}, {Qu{\'e}ruel}, {Jannet}, {Fergeau}, {Brochot}, {Cassam-Chenai},
  {Dudok de Wit}, {Timofeeva}, {Vincent}, {Agrapart}, {Delory}, {Turin},
  {Jeandet}, {Leroy}, {Pellion}, {Bouzid}, {Katra}, {Piberne}, {Recart},
  {Santol{\'\i}k}, {Kolma{\v{s}}ov{\'a}}, {Krupa{\v{r}}},
  {Krupa{\v{r}}ov{\'a}}, {P{\'\i}{\v{s}}a}, {Uhl{\'\i}{\v{r}}}, {L{\'a}n},
  {Ba{\v{s}}e}, {Ahl{\`e}n}, {Andr{\'e}}, {Bylander}, {Cripps}, {Cully},
  {Eriksson}, {Jansson}, {Johansson}, {Karlsson}, {Puccio},
  {B{\v{r}}{\'\i}nek}, {{\"O}ttacher}, {Panchenko}, {Berthomier}, {Goetz},
  {Hellinger}, {Horbury}, {Issautier}, {Kontar}, {Krucker}, {Le Contel},
  {Louarn}, {Martinovi{\'c}}, {Owen}, {Retino}, {Rodr{\'\i}guez-Pacheco},
  {Sahraoui}, {Wimmer-Schweingruber}, {Zaslavsky}, \&
  {Zouganelis}}]{2020A&A...642A..12M}
{Maksimovic}, M., {Bale}, S.~D., {Chust}, T., {et~al.} 2020, \aap, 642, A12

\bibitem[{{Mason}(2007)}]{2007SSRv..130..231M}
{Mason}, G.~M. 2007, Space Sci. Rev., 130, 231

\bibitem[{{Mason} {et~al.}(2000){Mason}, {Dwyer}, \&
  {Mazur}}]{2000ApJ...545L.157M}
{Mason}, G.~M., {Dwyer}, J.~R., \& {Mazur}, J.~E. 2000, Astrophys. J. Lett.,
  545, L157

\bibitem[{{Mason} {et~al.}(2020){Mason}, {Ho}, {Allen},
  {Rodr{\'\i}guez-Pacheco}, {Wimmer-Schweingruber}, \&
  {Bu{\v{c}}{\'\i}k}}]{2020A&AM}
{Mason}, G.~M., {Ho}, G.~C., {Allen}, R., {et~al.} 2020, \aap, in press

\bibitem[{{McComas} {et~al.}(1998){McComas}, {Bame}, {Barker}, {Feldman},
  {Phillips}, {Riley}, \& {Griffee}}]{1998SSRv...86..563M}
{McComas}, D.~J., {Bame}, S.~J., {Barker}, P., {et~al.} 1998, \ssr, 86, 563

\bibitem[{{M{\"u}ller} {et~al.}(2020){M{\"u}ller}, {St. Cyr}, {Zouganelis},
  {Gilbert}, {Marsden}, {Nieves-Chinchilla}, {Antonucci}, {Auch{\`e}re},
  {Berghmans}, {Horbury}, {Howard}, {Krucker}, {Maksimovic}, {Owen}, {Rochus},
  {Rodriguez-Pacheco}, {Romoli}, {Solanki}, {Bruno}, {Carlsson}, {Fludra},
  {Harra}, {Hassler}, {Livi}, {Louarn}, {Peter}, {Sch{\"u}hle}, {Teriaca}, {del
  Toro Iniesta}, {Wimmer-Schweingruber}, {Marsch}, {Velli}, {De Groof},
  {Walsh}, \& {Williams}}]{2020A&A...642A...1M}
{M{\"u}ller}, D., {St. Cyr}, O.~C., {Zouganelis}, I., {et~al.} 2020, \aap, 642,
  A1

\bibitem[{{Nitta} {et~al.}(2008){Nitta}, {Mason}, {Wiedenbeck}, {Cohen},
  {Krucker}, {Hannah}, {Shimojo}, \& {Shibata}}]{2008ApJ...675L.125N}
{Nitta}, N.~V., {Mason}, G.~M., {Wiedenbeck}, M.~E., {et~al.} 2008, Astrophys.
  J. Lett., 675, L125

\bibitem[{{Nitta} {et~al.}(2006){Nitta}, {Reames}, {De Rosa}, {Liu}, {Yashiro},
  \& {Gopalswamy}}]{2006ApJ...650..438N}
{Nitta}, N.~V., {Reames}, D.~V., {De Rosa}, M.~L., {et~al.} 2006, Astrophys.
  J., 650, 438

\bibitem[{{Owen} {et~al.}(2020){Owen}, {Bruno}, {Livi}, {Louarn}, {Al Janabi},
  {Allegrini}, {Amoros}, {Baruah}, {Barthe}, {Berthomier}, {Bordon},
  {Brockley-Blatt}, {Brysbaert}, {Capuano}, {Collier}, {DeMarco}, {Fedorov},
  {Ford}, {Fortunato}, {Fratter}, {Galvin}, {Hancock}, {Heirtzler}, {Kataria},
  {Kistler}, {Lepri}, {Lewis}, {Loeffler}, {Marty}, {Mathon}, {Mayall}, {Mele},
  {Ogasawara}, {Orlandi}, {Pacros}, {Penou}, {Persyn}, {Petiot}, {Phillips},
  {P{\v{r}}ech}, {Raines}, {Reden}, {Rouillard}, {Rousseau}, {Rubiella},
  {Seran}, {Spencer}, {Thomas}, {Trevino}, {Verscharen}, {Wurz}, {Alapide},
  {Amoruso}, {Andr{\'e}}, {Anekallu}, {Arciuli}, {Arnett}, {Ascolese},
  {Bancroft}, {Bland}, {Brysch}, {Calvanese}, {Castronuovo},
  {{\v{C}}erm{\'a}k}, {Chornay}, {Clemens}, {Coker}, {Collinson}, {D'Amicis},
  {Dandouras}, {Darnley}, {Davies}, {Davison}, {De Los Santos}, {Devoto},
  {Dirks}, {Edlund}, {Fazakerley}, {Ferris}, {Frost}, {Fruit}, {Garat},
  {G{\'e}not}, {Gibson}, {Gilbert}, {de Giosa}, {Gradone}, {Hailey}, {Horbury},
  {Hunt}, {Jacquey}, {Johnson}, {Lavraud}, {Lawrenson}, {Leblanc}, {Lockhart},
  {Maksimovic}, {Malpus}, {Marcucci}, {Mazelle}, {Monti}, {Myers}, {Nguyen},
  {Rodriguez-Pacheco}, {Phillips}, {Popecki}, {Rees}, {Rogacki}, {Ruane},
  {Rust}, {Salatti}, {Sauvaud}, {Stakhiv}, {Stange}, {Stubbs}, {Taylor},
  {Techer}, {Terrier}, {Thibodeaux}, {Urdiales}, {Varsani}, {Walsh}, {Watson},
  {Wheeler}, {Willis}, {Wimmer-Schweingruber}, {Winter}, {Yardley}, \&
  {Zouganelis}}]{2020A&A...642A..16O}
{Owen}, C.~J., {Bruno}, R., {Livi}, S., {et~al.} 2020, \aap, 642, A16

\bibitem[{{Reames}(2002)}]{2002ApJ...571L..63R}
{Reames}, D.~V. 2002, Astrophys. J. Lett., 571, L63

\bibitem[{{Reames}(2021)}]{2021LNP...978.....R}
{Reames}, D.~V. 2021, {Solar Energetic Particles}, Vol. 978 (Springer, Cham.)

\bibitem[{{Reames} \& {Stone}(1986)}]{1986ApJ...308..902R}
{Reames}, D.~V. \& {Stone}, R.~G. 1986, Astrophys. J., 308, 902

\bibitem[{{Rochus} {et~al.}(2020){Rochus}, {Auch{\`e}re}, {Berghmans}, {Harra},
  {Schmutz}, {Sch{\"u}hle}, {Addison}, {Appourchaux}, {Aznar Cuadrado},
  {Baker}, {Barbay}, {Bates}, {BenMoussa}, {Bergmann}, {Beurthe}, {Borgo},
  {Bonte}, {Bouzit}, {Bradley}, {B{\"u}chel}, {Buchlin}, {B{\"u}chner},
  {Cab{\'e}}, {Cadiergues}, {Chaigneau}, {Chares}, {Choque Cortez}, {Coker},
  {Condamin}, {Coumar}, {Curdt}, {Cutler}, {Davies}, {Davison}, {Defise}, {Del
  Zanna}, {Delmotte}, {Delouille}, {Dolla}, {Dumesnil}, {D{\"u}rig}, {Enge},
  {Fran{\c{c}}ois}, {Fourmond}, {Gillis}, {Giordanengo}, {Gissot}, {Green},
  {Guerreiro}, {Guilbaud}, {Gyo}, {Haberreiter}, {Hafiz}, {Hailey}, {Halain},
  {Hansotte}, {Hecquet}, {Heerlein}, {Hellin}, {Hemsley}, {Hermans}, {Hervier},
  {Hochedez}, {Houbrechts}, {Ihsan}, {Jacques}, {J{\'e}r{\^o}me}, {Jones},
  {Kahle}, {Kennedy}, {Klaproth}, {Kolleck}, {Koller}, {Kotsialos},
  {Kraaikamp}, {Langer}, {Lawrenson}, {Le Clech'}, {Lenaerts}, {Liebecq},
  {Linder}, {Long}, {Mampaey}, {Markiewicz-Innes}, {Marquet}, {Marsch},
  {Matthews}, {Mazy}, {Mazzoli}, {Meining}, {Meltchakov}, {Mercier}, {Meyer},
  {Monecke}, {Monfort}, {Morinaud}, {Moron}, {Mountney}, {M{\"u}ller},
  {Nicula}, {Parenti}, {Peter}, {Pfiffner}, {Philippon}, {Phillips},
  {Plesseria}, {Pylyser}, {Rabecki}, {Ravet-Krill}, {Rebellato}, {Renotte},
  {Rodriguez}, {Roose}, {Rosin}, {Rossi}, {Roth}, {Rouesnel}, {Roulliay},
  {Rousseau}, {Ruane}, {Scanlan}, {Schlatter}, {Seaton}, {Silliman}, {Smit},
  {Smith}, {Solanki}, {Spescha}, {Spencer}, {Stegen}, {Stockman}, {Szwec},
  {Tamiatto}, {Tandy}, {Teriaca}, {Theobald}, {Tychon}, {van Driel-Gesztelyi},
  {Verbeeck}, {Vial}, {Werner}, {West}, {Westwood}, {Wiegelmann}, {Willis},
  {Winter}, {Zerr}, {Zhang}, \& {Zhukov}}]{2020A&A...642A...8R}
{Rochus}, P., {Auch{\`e}re}, F., {Berghmans}, D., {et~al.} 2020, \aap, 642, A8

\bibitem[{{Rodr{\'\i}guez-Pacheco} {et~al.}(2020){Rodr{\'\i}guez-Pacheco},
  {Wimmer-Schweingruber}, {Mason}, {Ho}, {S{\'a}nchez-Prieto}, {Prieto},
  {Mart{\'\i}n}, {Seifert}, {Andrews}, {Kulkarni}, {Panitzsch}, {Boden},
  {B{\"o}ttcher}, {Cernuda}, {Elftmann}, {Espinosa Lara}, {G{\'o}mez-Herrero},
  {Terasa}, {Almena}, {Begley}, {B{\"o}hm}, {Blanco}, {Boogaerts}, {Carrasco},
  {Castillo}, {da Silva Fari{\~n}a}, {de Manuel Gonz{\'a}lez}, {Drews},
  {Dupont}, {Eldrum}, {Gordillo}, {Guti{\'e}rrez}, {Haggerty}, {Hayes},
  {Heber}, {Hill}, {J{\"u}ngling}, {Kerem}, {Knierim}, {K{\"o}hler}, {Kolbe},
  {Kulemzin}, {Lario}, {Lees}, {Liang}, {Mart{\'\i}nez Hell{\'\i}n}, {Meziat},
  {Montalvo}, {Nelson}, {Parra}, {Paspirgilis}, {Ravanbakhsh}, {Richards},
  {Rodr{\'\i}guez-Polo}, {Russu}, {S{\'a}nchez}, {Schlemm}, {Schuster},
  {Seimetz}, {Steinhagen}, {Tammen}, {Tyagi}, {Varela}, {Yedla}, {Yu},
  {Agueda}, {Aran}, {Horbury}, {Klecker}, {Klein}, {Kontar}, {Krucker},
  {Maksimovic}, {Malandraki}, {Owen}, {Pacheco}, {Sanahuja}, {Vainio},
  {Connell}, {Dalla}, {Dr{\"o}ge}, {Gevin}, {Gopalswamy}, {Kartavykh},
  {Kudela}, {Limousin}, {Makela}, {Mann}, {{\"O}nel}, {Posner}, {Ryan},
  {Soucek}, {Hofmeister}, {Vilmer}, {Walsh}, {Wang}, {Wiedenbeck}, {Wirth}, \&
  {Zong}}]{2020A&A...642A...7R}
{Rodr{\'\i}guez-Pacheco}, J., {Wimmer-Schweingruber}, R.~F., {Mason}, G.~M.,
  {et~al.} 2020, \aap, 642, A7

\bibitem[{{Scherrer} {et~al.}(2012){Scherrer}, {Schou}, {Bush}, {Kosovichev},
  {Bogart}, {Hoeksema}, {Liu}, {Duvall}, {Zhao}, {Title}, {Schrijver},
  {Tarbell}, \& {Tomczyk}}]{2012SoPh..275..207S}
{Scherrer}, P.~H., {Schou}, J., {Bush}, R.~I., {et~al.} 2012, \solphys, 275,
  207

\bibitem[{{Wang} {et~al.}(2016){Wang}, {Krucker}, {Mason}, {Lin}, \&
  {Li}}]{2016A&A...585A.119W}
{Wang}, L., {Krucker}, S., {Mason}, G.~M., {Lin}, R.~P., \& {Li}, G. 2016,
  Astron. Astrophys., 585, A119

\bibitem[{{Wang} {et~al.}(2006){Wang}, {Pick}, \&
  {Mason}}]{2006ApJ...639..495W}
{Wang}, Y.~M., {Pick}, M., \& {Mason}, G.~M. 2006, Astrophys. J., 639, 495

\bibitem[{{Wimmer-Schweingruber} {et~al.}(2021){Wimmer-Schweingruber},
  {Janitzek}, {Pacheco}, {Cernuda}, {Espinosa Lara}, \&
  {G{\'o}mez-Herrero}}]{2020A&AW}
{Wimmer-Schweingruber}, R.~F., {Janitzek}, N.~P., {Pacheco}, D., {et~al.} 2021,
  \aap, submitted

\end{thebibliography}
%

\begin{appendix} \label{app}
\section{EUV images of the solar sources}

\begin{figure}
   \centering
   \includegraphics[width=9cm]{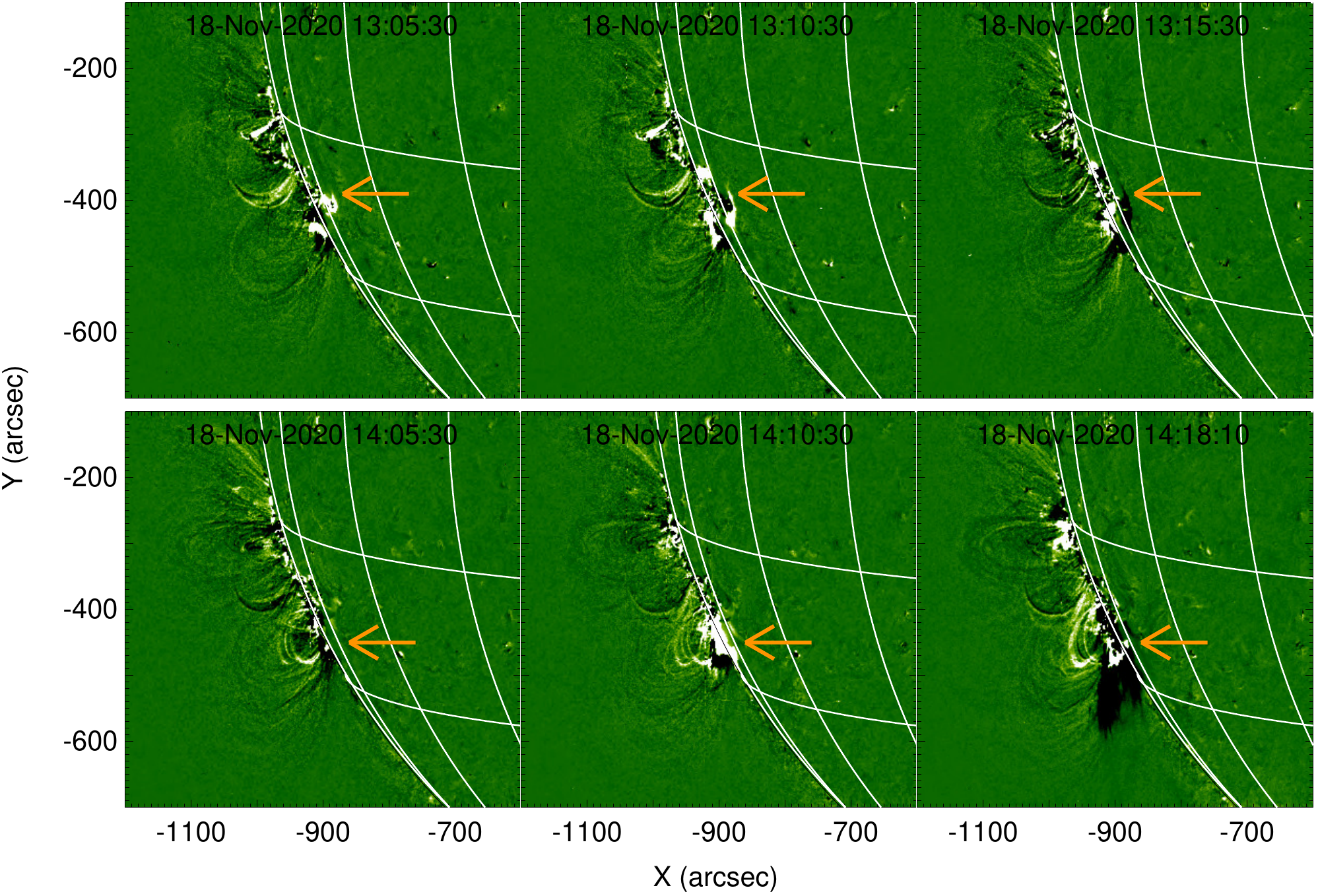}
   \caption{Same as Fig.~\ref{euv1} but for injection \#4. The top row corresponds to the 1st, and the bottom to the 2nd type III burst.}
              \label{euv2}%
    \end{figure}

\begin{figure}
  \centering
    \includegraphics[width=9cm]{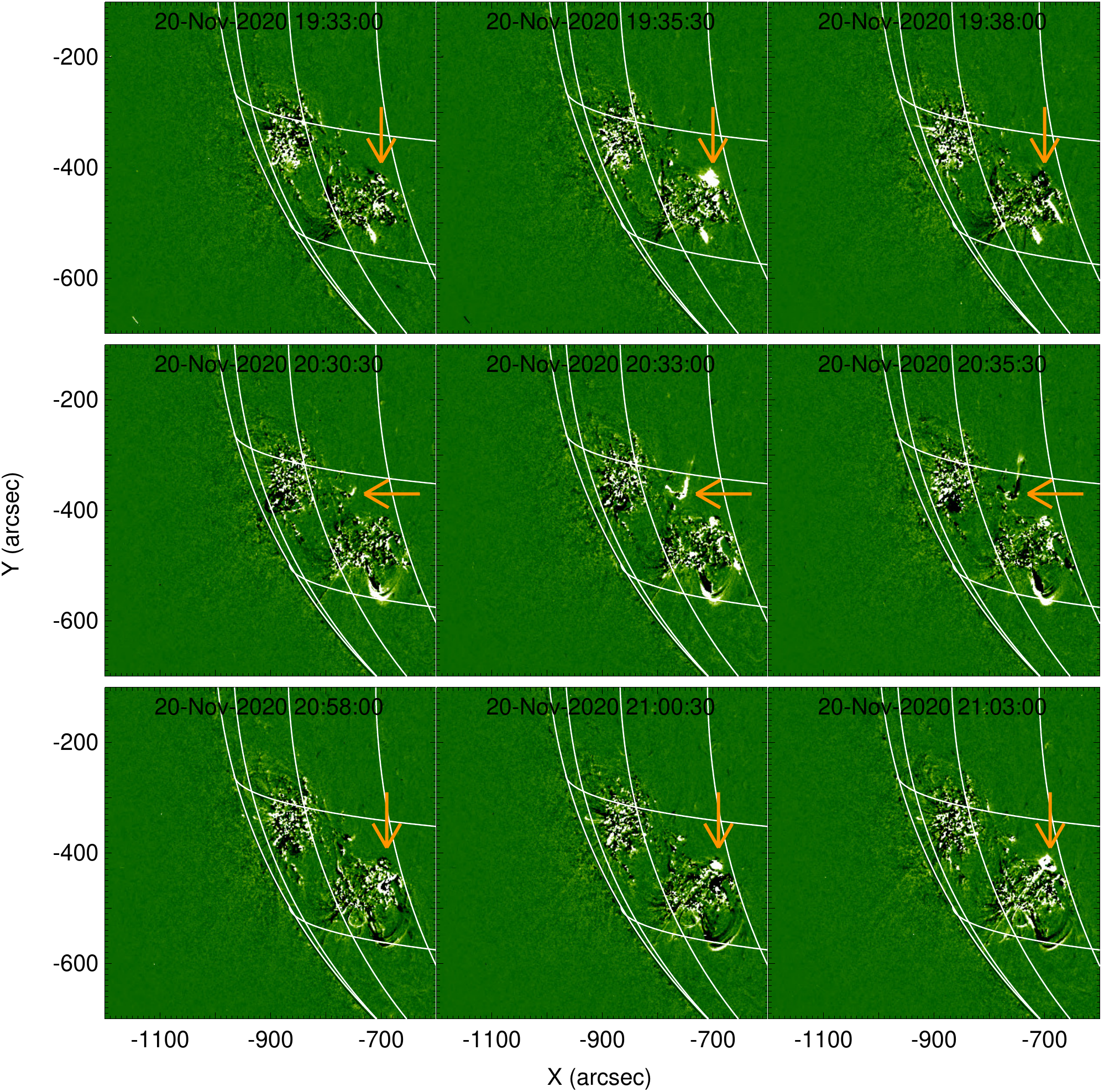}
   \caption{Same as Fig.~\ref{euv1} but for injection \#5. The top row corresponds to the 1st, middle to the 2nd, and bottom to the 3rd type III burst.}
              \label{euv3}%
    \end{figure}

\begin{figure}
  \centering
  \includegraphics[width=9cm]{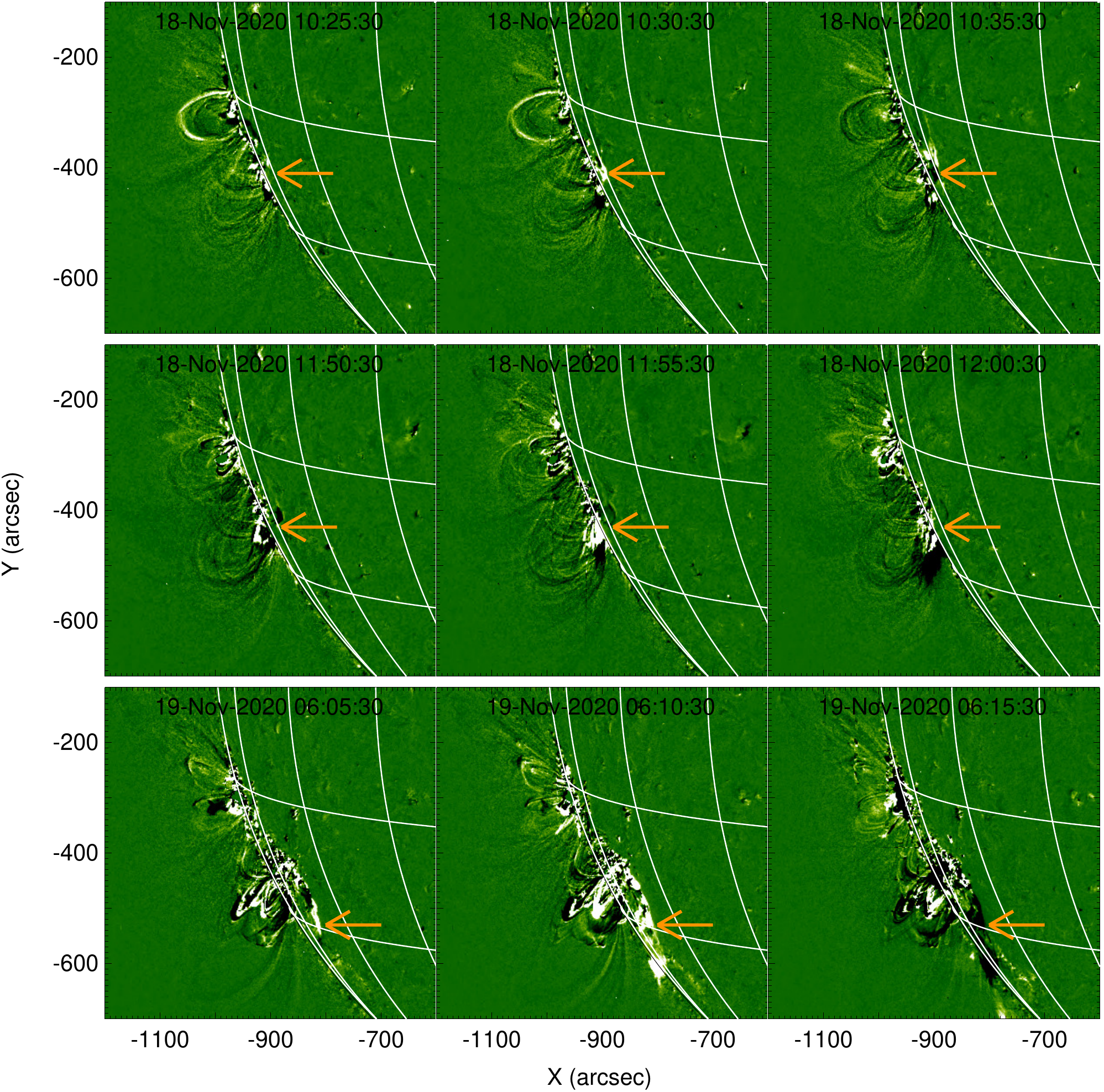}
   \caption{Same as Fig.~\ref{euv1} but for type III bursts that occurred during the decay phase of the 1st (top and middle panels) and the 2nd ion intensity increase (bottom panel).}
              \label{euv4}%
    \end{figure}

The solar sources associated with injections \#4, \#5, and the electron events/type III bursts that occurred during decay phase of the 1st and 2nd ion intensity increases are shown in Fig.~\ref{euv2}, Fig.~\ref{euv3}, and Fig.~\ref{euv4}, respectively.

\end{appendix}
\end{document}